\documentstyle[preprint, aps]{revtex}
\begin{document}
\title{Caloric curve in fragmentation}
\author{A. Strachan and C. O. Dorso}
\address{Departamento de F\'\i sica, Facultad de Ciencias Exactas y Naturales,\\
Universidad de Buenos Aires, Pabell\'on 1, Ciudad Universitaria, Nu\~nez\\
1428, Buenos Aires, Argentina.}
\maketitle

\begin{abstract}
We study phase transitions of small two dimensional Lennard Jones drops via
microcanonical molecular dynamics in a broad energy range. We found that the
caloric curve can be extended to high energies to comprehend high
evaporation rates and highly non-equilibrium phenomena such as
multifragmentation. Multifragmentation appears as a constant temperature
region in the caloric curve like the solid-liquid phase transition.
\end{abstract}


The problem of phase transitions in small many body systems has attracted
much attention in different areas of physics during the last two decades and
remains of great importance. Among these processes, and particularly
important in nuclear physics we can mention multifragmentation. In this
phenomena a very excited small system undergoes a highly non equilibrium 
process, in 
which a collective expansion develops and the initial drop breaks up in 
several intermediate mass fragments. These fragmentation processes are 
studied in many branches of physics, for example fragmentation in 
intermediate mass heavy ion
collisions \cite{nuclear}, drop formation in fluids flowing through
nozzles, lithium jets bombarded with neutrons in inertial confinement
reactors \cite{blink}, fragmentation in cluster deposition for the synthesis
of nanostructured surfaces \cite{bromann}. In spite of the great efforts
done is order to study these phenomena many questions remain unsolved, the
mechanisms that make the excited drop break into several fragments and
the possible connection of these processes with the liquid-vapor phase 
transition are not clear at present \cite{cod-pbc}. In the nuclear 
multifragmentation case, we would like to mention the current efforts
towards the determination of the caloric curve by the
ALADIN collaboration \cite{pocho}, and the calculations of different
critical exponents by the EOS collaboration \cite{eos}.

When dealing with infinite systems one usually thinks on solid-liquid and
liquid-vapor phase transitions. It is well known that small finite systems 
display solid-like to liquid-like transition, which can be thought of as the 
rounding-off of the first order phase transition of the infinite case due to
finite size effects. On the other hand the process of fragmentation
described above is a finite system thing, and does not appear in the 
thermodynamic limit.

The analysis of  the solid-like to liquid-like phase transition of very
small systems, and its connection to the macroscopic solid-liquid phase
transition has relied strongly on computer simulations. One of the most
relevant quantities regarding this phenomena is the caloric curve ($T(E)$), 
which displays a rather constant region at the phase transition. As
expected for small systems the behavior depends on the thermodynamic
ensemble used in the simulation \cite{labastie90}. Different simulation
methods were used, for example constant temperature molecular dynamics \cite
{berry94}, constant ($T,P,N$) Monte Carlo \cite{berry92} and microcanonical
molecular dynamics \cite{chakra}. The upper limit of the range of energies
studied in these works is chosen so as to avoid dealing with evaporation.
Although this phase transition is not relevant for the nuclear case, we
will show that its study can be of great importance in order to understand
fragmentation in classical many body systems.

Consider a small isolated drop in an infinite volume; according to the
energy of the drop we classify schematically its behavior in three different
cases. a) Low energy drops, where evaporation does not occur, at least
during the time scales considered, this case contains the solid-like to
liquid-like phase transition. For higher energy we will get b) evaporating
drops, where few particles evaporate during the experiment. It is worth
mentioning at this point that this kind of experiment, as any of this kind
with free boundary conditions, is to be thought as equivalent to a
time-scale constraint in the experiment conducted at zero pressure \cite
{berry92}. For even higher energy c) we enter the case of
multifragmentation. The initial state of this experiment is a hot and
compressed drop which expands and breaks up in several fragments. It was
shown in \cite{times} that some of the initial thermal excitation is
transformed, via inter-particle collisions, into a collective radial
expansion, and consequently the system cools down. The time scale considered
in the present work is related to the one of fragment formation for the
multifragmentation experiments \cite{times}.

In this communication we study the behavior of two dimensional drops in a wide
energy range, embodying the solid-like to liquid-like transition,
evaporating drops, and the phenomena of fragmentation. We first describe the
microscopic model that we use to simulate the atomistic system. We then
present our multifragmentation computer experiments and the information that
can be extracted from them, followed by our results for equilibrated drops,
which contains the solid-liquid phase transition. The connection of the two
phenomena is then shown.

We simulate the atomistic two dimensional system formed by $N=100$ Lennard
Jones particles interacting via the two body potential:

\begin{equation}
V(r)=4\epsilon \left[ \left( \frac \sigma r\right) ^{12}-\left( \frac \sigma 
r\right) ^6-\left( \frac \sigma {r_c}\right) ^{12}+\left( \frac \sigma {r_c}%
\right) ^6\right] , 
\end{equation}

where $r_c$ is the cutoff radius, the potential is taken as zero for $r\geq
r_c$. We took $r_c=3\sigma $. The unit of time and energy are $t_0=\sqrt{%
\sigma ^2m/48\epsilon }$ and $\epsilon $ respectively. The time step that we
used for the integration of the equation of motion was fixed to $%
t_{int}=0.0025t_0$, in this way the energy and total angular momentum is
conserved better than $0.01\%$.

For the study of multifragmentation we analyzed the time evolution of
compressed and hot drops. The initial configurations are obtained by cutting
a circular drop from a thermalized two dimensional Lennard Jones periodic
system. These initial configurations are macroscopically characterized by
the energy $E$, the density $\rho $ (of the periodic system) and the number
of particles which is fixed to $N=100$, see \cite{times} for details.

Defining the temperature in these experiments is not an easy task. The
process of fragmentation can be divided in two different stages. During the
first stage of the evolution ``flux and fragment formation stage''
a collective expansion builds up, and the asymptotic clusters
are formed in phase space, according to the ECFM (early cluster
formation model) \cite{dorso93,times}.
During the second stage ``fragment emission stage'' the clusters, already 
formed in phase space \cite{times}, spread in coordinate space, i.e. they
are emitted. 
It is worth mentioning at this point that the ECFM allows us to
preciselly identify the break-up or fragment formation time.

In order to study the cooling of the system while the expansion builds up,
we can define for the initial dense stages a local temperature, which
depends on time and position, in the following way. We define the local
temperature $T_l({\bf {r}},t)$ as the velocity fluctuations around the mean
radial expansion, which is position dependent (the outer regions expand
faster than the inner ones). In other words we make the conjecture that
our system is in local equilibrium \cite{huang}, i.e. the velocity
distribution of all the particles of our system depend on position and satisfy:
\begin{equation}
	f({\bf v};{\bf r}) = \rho({\bf r}) 
		\left( \frac{m \beta({\bf r})}{2 \pi} \right) \,
		 e^{ \beta({\bf r}) \frac{m}{2} 
		({\bf v} - {\bf v}_{rad}({\bf r}))^2 }
\end{equation}
where  $\rho({\bf r})$ and  $\beta({\bf r})$ are the density and inverse
of the temperature respectively; ${\bf v}_{rad}({\bf r})$ is the collective
velocity which in our case, due to the symmetry of the problem, is in 
the radial direction.
The actual calculation is done in the following way:
we divide the system in ten concentric circular regions, centered in the
c.m. of the system, and we calculate the mean radial velocity of each
circular ring. The temperature of each circular region is defined as the
velocity fluctuations around the meassured collective motion (mean
radial velocity)  of the corresponding
ring, \cite{times}. The total kinetic energy is the sum of the one 
corresponding to the expansion and that associated to the local temperature.
The general picture obtained is the following: the
process involves a fast cooling while the expansion builds up, and after a
time $\tau (E)$ the temperature remains almost constant in time
\cite{times}. Furthermore
the temperature of the central regions reach approximately the same value and
the radial and transverse velocity fluctuations (or local temperatures)
are very similar. This last fact shows that the system reaches some
degree of local thermalization 
\cite{times}, in this way we see that our conjecture
of local equilibrium is justified. 

We can then define
the temperature of the system as a whole, $\left\langle T_l(t)\right\rangle $,
as the average over the temperatures of the inner regions so as to include
approximately $80$ particles; the outer rings are disregarded because they
are populated with only few free particles and very small aggregates. 
This definition is only valid as
long as the system is still dense. 

In Fig. 1 we show $\left\langle
T_l\right\rangle $ as a function of time, for different initial energies and
densities, see caption for details. It can be seen the fast cooling during
the first stage of the evolution. 
The initial conditions shown in Fig. 1 represent a very wide range of 
energies as reffers to the fragmentation pattern (the initial temperature 
range is $1\epsilon \leq \left\langle T_l(0)\right\rangle \leq 4\epsilon $). 
For $E = 0.3 \epsilon$ the system breaks into several small fragments 
and the mass spectrum is exponentially decaying, while
for $E = -1.1 \epsilon$ the mass spectrum is U shaped, at this low
energy an important part of the system remains bound, see \cite{times}.
From Fig.1 we get the important result that while the collective expansion
builds up the system cools down until the local temperature reaches a given
value, which lies between $0.35\epsilon $ and $0.4\epsilon $, independent of
the initial conditions, i.e. density and energy. From this time on, the
system continues its expansion at a constant radial velocity and breaks-up. 

Asymptotically the system is formed by a set of non interacting clusters
that are moving away from each other, it is clear then that we can not speak
of the temperature of the system as a whole.
In this case, when fragments are already formed, the total kinetic energy of
the system can be expressed in the following way

\begin{eqnarray}
K_{total} =\sum_{clusters}\frac 1{2m_c}{\bf P}_c^2+ \sum_{clusters}K_i^{c.m.} 
+ \sum_{clusters}\frac{I_c\omega _c^2}2
\end{eqnarray}

where $P_c$ is the momentum of the center o mass of cluster $c$ of total mass 
$m_c$ and the last term is the total rotational kinetic
energy of the clusters. $K_i^{c.m.}$ is, then, the kinetic energy of
particles belonging to cluster $c$ once the total linear and angular 
momentum are removed.

We now focus on the temperature of the asymptotic fragments $T_f$, which is
given by : 
\begin{equation}
T_f(i)=\frac 1{n_i-3/2}K_i^{c.m.},
\end{equation}
where $n_i$ denotes the mass number of cluster $i$. In Fig. 2 we show the
average temperature of the asymptotic fragments of our molecular dynamics
experiments as a function of their mass, for four different initial
conditions: $E=-1.1\epsilon $ , $\rho =0.75\sigma ^{-2}$ (full circles), $%
E=-0.55\epsilon $, $\rho =0.75\sigma ^{-2}$ (empty circles), $E=0.8\epsilon $%
, $\rho =0.75\sigma ^{-2}$ (diamonds) $E=-0.3\epsilon $ , $\rho =0.85\sigma
^{-2}$ (triangles). No less than $100$ events for each initial condition
were averaged. Again the cases shown in 
Fig. 2 represent a wide range of excitation as refers to the behavior of the
system. It can be seen that for $E=0.8\epsilon $ there are no clusters of
mass bigger that $20$, at this high energy the system breaks in several
small mass fragments. On the other hand, for $E=-1.1\epsilon $, there are
very few intermediate mass fragments which comes from a U shaped mass
spectra, the mass spectra for most of the cases are shown in \cite{times}. 
Fig. 2 contains a remarkable result: despite the wide range of
excitations shown, the temperature of clusters with a given mass number $n$
is fixed no matter the initial condition of the system.

It is important to note that the temperature of the intermediate and 
big mass clusters is 
equal to the local temperature (velocity fluctuations around the mean
radial velocity as defined above) at
the break up time (see Fig. 1). In order to gain a better comprehension of
the multifragmentation process, it is important to understand what is so
particular about this value of temperature, so we studied our $N=100$ LJ
particles system for lower energies, when fragmentation does not occur.

The study of the low energy regimes of small systems usually involve the
calculation of caloric curves ($T(E)$), which might give us some information
about the meaning of the value of temperature discussed in the last
paragraph. The usual caloric curves show three regions: a) solid region in
which $T$ increases steadily leading to constant specific heat, b) the
solid-like to liquid-like transition in which $T$ is rather constant, c)
liquid region in which, again, $T$ increases \cite{labastie90}. It is clear
that the liquid-like state cannot be heated without limits, so another
plateau is to be expected.

In order to study this we calculated the caloric curve of our two
dimensional Lennard Jones drops. We imposed free boundary conditions to our
system. In this way we reproduce the conditions at which multifragmentation
experiments take place, where the system is not confined. The initial
configuration was constructed in the same way as for multifragmentation, but
this time the energy and the density were close to those of the ground state
of the drop. When the drop is evolved the particles display approximately a
triangular lattice; the system does not explode, nor does it evaporate
particles. In order to study the caloric curve ($T(E)$), after performing
the isoenergetic evolution of the drop the velocities of the particles of
the last configuration are re-scaled in order to increase the energy in a
constant amount, and a new isoenergetic evolution is performed. This
procedure is repeated in order to cover the desired range of energies. If
the system, in its evolution, evaporates one or more particles they are
disregarded for the next evolution, at a slightly higher energy; only the
particles remaining in the drop are considered in the velocity re-scaling
after their center of mass velocity and angular momentum are removed. With
our procedure the system will loose particles, due to evaporation, as it is
heated. For our calculations we only considered configuration in which the
number of particles in the remaining drop was $90\leq N\leq 100$. This
variation in the mass of the system is not a serious problem since we found
that the temperature and other macroscopic quantities of our system do not
change when few particles are evaporated.

The temperature for a given energy is given by the well known relation for
2-D:

\begin{equation}
T=\left\langle \frac 1{N(t)-3/2}\sum_{i=1}^{N(t)}\frac 12m{\bf v}%
_i^2\right\rangle _t, 
\end{equation}
where $\left\langle {}\right\rangle _t$ means time average and $N(t)$ is the
number of particles which at time $t$ have not evaporated. The time average
is done once the system has reached equilibrium, the first $2\times 10^4$
time steps accomplish this thermalization. The total time considered for the 
average is $150t_0$ which
is approximately the time of the multifragmentation experiments. It is worth
mentioning that this last definition of temperature is equivalent to those
used in multifragmentation because for these low energy cases the collective
expansion can be disregarded, no more than three particles are evaporated
for a given energy in the period of time considered.

In this way we studied the range of energy $-2.8\epsilon $ to $-1.4\epsilon $
of the caloric curve, see the circles in Fig. 3. Three different regions can
be identified. I) Solid region in which $T$ increases steadily, II)
solid-liquid transition, in which $T$ is rather constant, and III) liquid
region in which $T$ increases steadily again and then presents the beginning
of a high energy plateau ($-1.8\epsilon \leq E\leq -1.4\epsilon $). Although
the plateau denoting the solid-like to liquid-like phase transition is very
small, the transition as one increases the energy is also denoted by
particles being popped out of the surface and occupying the adjacent outer
layer \cite{berry94} and by the sudden increase of the root mean square bond
length fluctuation (Lindemann index) above $0.1$ \cite{chakra} (not shown in
the figure for clarity). The solid-like to liquid-like phase transition in
small clusters has been extensively studied, see references in the first
paragraph, and although this is, to the best of our knowledge, the first
time that this is done in two dimensional systems, we are not interested in
a detailed description of the solid-liquid phase transition but in the
general behavior of the caloric curve and its connection with
multifragmentation.

In the high energy part of region III the evaporation is quite important and
it is very difficult to attain metastable liquid-like configurations for
long enough times if the energy is increased beyond $-1.4\epsilon $, this
fact makes the calculation of the caloric curve by this procedure
meaningless beyond this energy. This is because for higher energies we enter
the multifragmentation regime, described above. Remember that in this case
the initial state of the system is out of equilibrium and characterized by
energy and the density and the main de-excitation channel is the creation of
a collective expansion, the system does not simply evaporate particles but
expands and breaks into several fragments. The values of $T$ for energies
higher that $-1.4\epsilon $ in Fig. 3 were obtained from the results shown
in figures 1 and 2. The temperature corresponding to a given energy is
calculated in two ways. a) squares show the local temperature at break-up
time, $\left\langle T_l\right\rangle $
averaged over a period of time of approximately $30t_0$ centered at the
break up time (the break-up times can be found in \cite{times}). 
This is basically the
constant value of the local temperature shown in Fig. 2. 
b) The diamonds
show the average cluster temperature (we only considered fragments of mass
number greater that $15$). See caption for the initial densities. It can be
seen that the asymptotic temperature of the clusters is quite similar to the
local temperature at the break-up time, moreover we get the important result
that the low energy part of the caloric curve joins smoothly the results of
the multifragmentation experiments. The full caloric curve features two
plateaux, the low energy one is related to the solid-like to liquid-like
transition, and the high energy ones denotes the evaporating liquid and
multifragmentation. This behavior of the caloric curve is in agreement
with very recent experimental results for the nuclear caloric curve
\cite{hauger,serfling}. On the other hand the experimental result
of \cite{pocho} show a caloric curve featuring a plateau followed
by a rise of temperature for high excitation energy, which was denoted 
``gas phase''. This ``gas phase'' behavior was also found in computer
simulations of nuclear fragmentation by Bondorf et. al. \cite{bondorf}. 
In this last work the temperature is related 
to the c.m. velocity of the asymptotic clusters, making strong
assumptions regarding the flow velocity distribution. Let us stress
that our approach makes no assumptions about the radial flow, in
this way we make sure that the collective motion is correctly
calculated and subtracted in order to compute the temperature
of the expanding system.
We do not find any increase of temperature in the fragmentation
region, and within our picture of fragmentation, explained in the
next paragraph, no ``gas phase'' is to expected in fragmentation
experiments. Its presence in \cite{bondorf} might be due to
the fact that the collective motion is not taken into account
properly. 

From the results presented in Fig. 3, the connection between
multifragmentation and the liquid-gas phase transition appears. From the
region of highly evaporating liquid of Fig. 3 it can be seen that it is not
possible to increase the temperature of a liquid drop beyond some limit
value which for our system is $\sim 0.37\epsilon $, we will call it the
limit temperature; as we increase the energy, the evaporation will become
more important but the system will not heat up. In our multifragmentation
experiments the initial state is artificially constructed hotter than the
limit case, in real systems this highly excited state is obtained via a
sudden input of energy to the system, for example by collisions. As the
system evolves it rapidly expands and cools down until the limit value for
temperature is achieved and the system needs no further relaxation, only
some evaporation will occur. If the energy of the initial condition is
increased, the radial collective velocity will grow and the system will
break into smaller fragments but the asymptotic temperature will not change,
in the same way as in the solid-liquid phase transition in which an input of
energy will not result in an increase of temperature but in melting. When
the system breaks, the clusters will be as hot as they can, which is
determined by the limit temperature, this is why the temperature of the
asymptotic clusters as a function of their mass is independent of the
initial conditions, (Fig. 2).

In conclusion we found that the microcanonical caloric curve can be extended
to high energies in such a way that it can describe the non equilibrium
process of multifragmentation, which appears as a constant temperature
region. This shows that although fragmentation is not a phase transition in
a strict sense, e.g. it does not show phase coexistence, it has some
properties that are characteristic of phase transitions. One of these is the
flat caloric curve, other being that the mass spectra is similar to those
found in the liquid-gas phase transition (see \cite{times}and references
therein). We also found that during the first stages of multifragmentation
the system transforms thermal motion into a more ordered collective
expansion until it reaches the limit temperature, independent of the initial
conditions, which is the maximum attainable temperature for a liquid drop.
This fact enable us to know in advance the velocity fluctuations over the
collective expansion when the fragmenting system breaks. This is the first
step towards a full characterization of the break up state in
multifragmentation, the other important ingredients being the radial
velocity field and the density fluctuations in phase space.

\acknowledgements
The authors gratefully acknowledge E. S. Hernandez for carefully reading
this manuscript. This work was done under partial financial support from the
University of Buenos Aires via grant EX-070, C. O. D. is member of the
Carrera del Investigador Cientifico CONICET- ARGENTINA, A. S. is a fellow of
the University of Buenos Aires.

\figure{FIG. 1. Local temperature as a function of time, for 
different initial conditions.
Full line denote $\rho=0.75(1/\sigma^2)$ and $E=-1.1\epsilon$, dotted line $%
\rho=0.75(1/\sigma^2)$ and $E=-0.55\epsilon$, dashed line $%
\rho=0.75(1/\sigma^2)$ and $E=0.3\epsilon$ and dashed-dotted line $%
\rho=0.85(1/\sigma^2)$ and $E=-0.3\epsilon$.}

\figure{FIG. 2. Temperature of the asymptotic clusters as a function of the
mass number, for different initial conditions. Full circles denote $%
\rho=0.75(1/\sigma^2)$ and $E=-1.1\epsilon$, empty circles $%
\rho=0.75(1/\sigma^2)$ and $E=-0.55\epsilon$, empty diamonds $%
\rho=0.75(1/\sigma^2)$ and $E=0.8\epsilon$ and empty triangles $%
\rho=0.85(1/\sigma^2)$ and $E=-0.3\epsilon$.}

\figure{FIG. 3. Caloric curve. Four regions can be identified: I solid-like,
II solid-like to liquid-like phase transition, III liquid-like and IV
multifragmentation. Circles denote results from the equilibrium experiments,
squares denote time average of the local temperature centered at the
break-up in fragmentation and diamonds show the mean temperature of
asymptotic clusters of mass number $>15$ in multifragmentation. The initial
density in the multifragmentation experiments is $\rho=0.75(1/\sigma^2)$ for
all energies but for $E=-0.3\epsilon$ and $E=-0.5\epsilon$ in which it is $%
\rho=0.85(1/\sigma^2)$ }

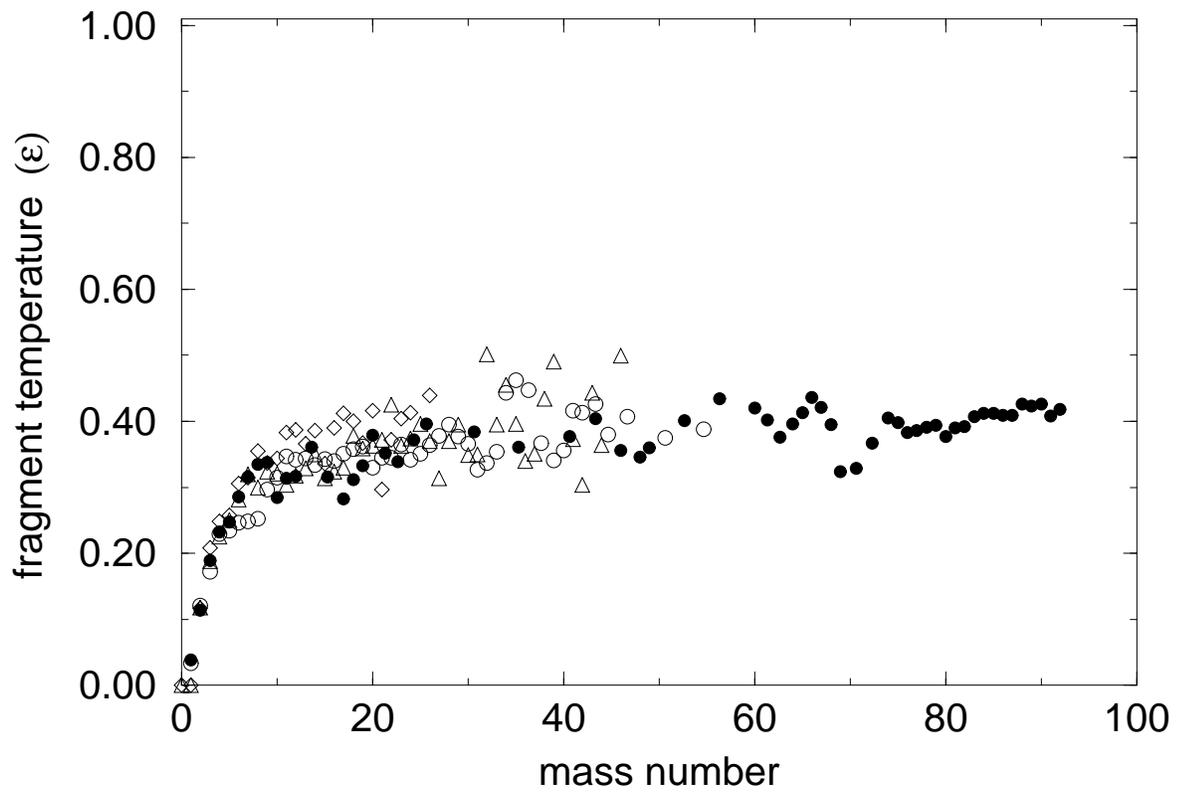
\end{document}